\documentclass[epsfig,a4paper,12pt,amsmath]{article}
\usepackage[dvips]{epsfig}
\usepackage{amsmath}

\large

\begin{document}

\title{\bf  Charge Asymmetry and Photon Energy Spectrum in the Decay $B_s \to l^+ l^- \gamma$}
\author{\bf Yusuf Din\c{c}er\footnote{e-mail:dincer@physik.rwth-aachen.de} \,  and Lalit M. Sehgal\footnote{e-mail:sehgal@physik.rwth-aachen.de} \\
   Institute of  Theoretical Physics, RWTH Aachen \\ D-52056 Aachen,  Germany}
\date{}
\maketitle


\begin{abstract}

We consider the structure-dependent amplitude of the decay $B_s \to l^+ l^- \gamma$ $(l=e,\mu)$ in a model based on the effective Hamiltonian for $b \bar{s} \to l^+ l^-$ 
containing the Wilson coefficients $C_7,C_9$ and $C_{10}$. The form factors characterising the matrix elements 
$\left< \gamma | \bar{s} \gamma_\mu (1 \mp \gamma_5) b | \bar{B}_s \right>$
and
$\left< \gamma | \bar{s} \sigma_{\mu\nu} (1 \mp \gamma_5) b | \bar{B}_s \right>$ 
are taken to have the universal form $f_V \approx f_A \approx f_T \approx f_{B_s} M_{B_s} R_s / (3 E_\gamma)$ suggested by recent work in QCD, where $R_s$ is a parameter
related to the light cone wave function of the $B_s$ meson. Simple expressions are obtained for the charge asymmetry $A(x_\gamma)$ and the photon energy spectrum 
$d \Gamma/ d x_\gamma (x_\gamma = 2 E_\gamma/M_{B_s})$. The decay rates are calculated in terms of the decay rate of $B_s \to \gamma \gamma$. 
The branching ratios are estimated to be $Br( B_s \to e^+ e^- \gamma) = 2.0 \times 10^{-8}$ and $Br(B_s \to \mu^+ \mu^- \gamma) = 1.2 \times 10^{-8}$, somewhat higher
than earlier estimates.

\end{abstract}


\newpage

\section{\bf Introduction}\label{introduction}

The rare decay $B_s \to l^+ l^- \gamma$ is of interest as a probe of the effective Hamiltonian for the transition $b \bar{s} \to l^+ l^-$, and as a testing ground for form 
factors describing the matrix elements $\left< \gamma | \bar{s} \gamma_\mu (1 \mp \gamma_5) b | \bar{B}_s \right>$
and
$\left< \gamma | \bar{s} i \sigma_{\mu\nu} (1 \mp \gamma_5) b | \bar{B}_s \right>$ \cite{aliev,geng}.
The branching ratio for $B_s \to l^+ l^- \gamma$ can be sizeable in comparison to the non-radiative process $B_s \to l^+ l^-$, since the chiral suppression of the latter is 
absent in the radiative transition. We will be concerned mainly with the structure-dependent part of the matrix element, since the correction due to bremsstrahlung from the 
external leptons is small and can be removed by eliminating the end-point region $s_{l^+ l^-} \approx M^2_{B_s}$.
(For related studies of radiative B decays, we refer to the papers in Ref. \cite{eilam}.)

Our objective is to calculate the decay spectrum of $B_s \to l^+ l^- \gamma$ using form factors suggested by recent work in QCD \cite{kor}. 
These form factors have the virtue of 
possessing a universal behaviour $1/E_\gamma$ for large $E_\gamma$, as well as a universal normalization. These features can be tested in measurements of 
$B^+ \to \mu^+ \nu \gamma$ and $B_s \to \gamma \gamma$. We derive simple formulae for the photon energy spectrum $d \Gamma/d x_\gamma$, $x_\gamma = 2 E_\gamma/m_{B_s}$, and 
the charge asymmetry $A(x_\gamma)$, defined as the difference in the probability of events with $E_+ > E_-$ and $E_+ < E_-$, $E_\pm$ being the $l^\pm$ energies. This
asymmetry is large over most of the $x_\gamma$ domain. Predictions are obtained for the branching ratios $Br( B_s \to e^+ e^- \gamma)$ and $Br( B_s \to \mu^+ \mu^- \gamma)$
which are somewhat higher than those estimated in previous literature \cite{aliev,geng}.


\section{\bf Matrix Element and Differential Decay Rate}\label{effective_hamiltonian}

The effective Hamiltonian for the interaction $b \bar{s} \to l^+ l^-$ has the standard form \cite{buch}
\begin{equation}\label{effechamil}
\begin{split}
{\cal H}_{\rm eff} 
&= \frac{\alpha G_F}{\sqrt{2} \pi} V_{tb} V^\star_{ts}  
\biggl\{
C^{\rm eff}_9        (\bar{s} \gamma_\mu P_L b)         \bar{l} \gamma_\mu l          \\
&+
C_{10}              (\bar{s} \gamma_\mu P_L b)           \bar{l} \gamma_\mu \gamma_5 l \\
&- 
2 \frac{C_7}{q^2} \bar{s} i \sigma_{\mu\nu} q^\nu (m_b P_R + m_s P_L) b \bar{l} \gamma_\mu l  
\biggr\}   
\end{split}
\end{equation}
where $P_{L,R} = (1 \mp \gamma_5)/2$ and $q$ is the sum of the $l^+$ and $l^-$ momenta. For the purpose of this paper, we will neglect the small $q^2-$dependent terms in 
$C^{\rm eff}_9$, arising from one-loop contributions of four-quark operators, as well as long-distance effects associated with $c \bar{c}$ resonances. The Wilson coefficients
in Eq.(\ref{effechamil}) will be taken to have the constant values
\begin{equation}
C_7 = -0.315 \; , \; C_9 = 4.334 \; , \; C_{10} = -4.624
\end{equation}

To obtain the amplitude for $B_s \to l^+ l^- \gamma$, one requires the matrix elements $\left< \gamma | \bar{s} \gamma_\mu (1 \mp \gamma_5) b | \bar{B}_s \right>$
and
$\left< \gamma | \bar{s} i \sigma_{\mu\nu} (1 \mp \gamma_5) b | \bar{B}_s \right>$. We parametrise these in the same way as in Ref. \cite{aliev,geng}
\begin{equation}\label{matrixelement}
\begin{split}
\left< \gamma(k) | \bar{s} \gamma_\mu b | \bar{B}_s (k+q) \right> 
&= 
e \, \epsilon_{\mu \nu \rho \sigma} \epsilon^{\star \nu} q^\rho k^\sigma f_V( q^2)/M_{B_s}  \\
\left< \gamma(k) | \bar{s} \gamma_\mu \gamma_5 b | \bar{B}_s (k+q) \right> 
&=
 -i e \biggl[ \epsilon_\mu^\star k \cdot q - \epsilon^\star \cdot q k_\mu \biggr] f_A(q^2)/M_{B_s} \\
\left< \gamma(k) | \bar{s} i \sigma_{\mu \nu} q^\nu b | \bar{B}_s (k+q) \right> 
&=
 - e \, \epsilon_{\mu \nu \rho \sigma} \epsilon^{\star \nu} q^\rho k^\sigma f_T(q^2) \\
\left< \gamma(k) | \bar{s} i \sigma_{\mu \nu} \gamma_5 q^\nu b | \bar{B}_s (k+q) \right> 
&=
 - i e \biggl[ \epsilon^\star_\mu k \cdot q - \epsilon^\star \cdot q k_\mu \biggr] f^\prime_T (q^2) 
\end{split}
\end{equation}
The form factors $f_V, f_A, f_T$ and $f^\prime_T$ are dimensionless, and related to those of Aliev et al \cite{aliev} by
$f_V = g/M_{B_s}, f_A = f/M_{B_s} , f_T = - g_1/M^2_{B_s}, f^\prime_T = - f_1/M^2_{B_s}$. The matrix element for $\bar{B}_s \to l^+ l^- \gamma$ can then be written as 
(neglecting terms of order $m_s/m_b)$
\begin{equation}
\begin{split}
{\cal M} (\bar{B}_s \to l^+ l^- \gamma) 
&=
\frac{\alpha G_F}{2 \sqrt{2} \pi} e V_{tb} V^\star_{ts} \frac{1}{M_{B_s}} \\
& 
\biggl[ \epsilon_{\mu \nu \rho \sigma} \epsilon^{\star \nu} q^\rho k^\sigma \biggl( A_1 \bar{l} \gamma^\mu l + A_2 \bar{l} \gamma^\mu \gamma_5 l \biggr) \\
&+ 
i \biggl( \epsilon^\star_\mu (k \cdot q) - (\epsilon^\star \cdot q) k_\mu \biggr) \biggl( B_1 \bar{l} \gamma^\mu l + B_2 \bar{l} \gamma^\mu \gamma_5 l \biggr) \biggr] 
\end{split}
\end{equation}
where 
\begin{equation}\label{a1etc}
\begin{split}
A_1 &= C_9 \, f_V  + 2 \, C_7 \, \frac{M^2_{B_s}}{q^2} \, f_T \\
A_2 &= C_{10} \, f_V      \\
B_1 &= C_9 \, f_A + 2 \, C_7 \frac{M^2_{B_s}}{q^2} \, f^\prime_T        \\
B_2 &= C_{10} \, f_A       
\end{split}
\end{equation}
(In the coefficient of $C_7$, we have approximated $m_b M_{B_s}$ by $M^2_{B_s}$). The Dalitz plot density in the energy variables $E_\pm$ is
\begin{equation}
\frac{d \Gamma}{d E_+ d E_-} = \frac{1}{256 \pi^3 M_{B_s}} \sum_{\rm spin} | {\cal M} |^2
\end{equation}
where \cite{aliev,geng,xiong}
\begin{equation}
\begin{split}
\sum_{\rm spin} | {\cal M} |^2 
&= 
 \biggl| \frac{\alpha G_F}{ \sqrt{2} \pi} V_{tb} V^\star_{ts} e \biggr|^2 \frac{1}{M^2_{B_s}} \\
& 
 \biggl\{ (|A_1|^2 + |B_1|^2) \biggl[  q^2 \{(p_+ \cdot k)^2 + (p_- \cdot k)^2 \} + 2m^2_l (q \cdot k)^2   \biggr]      \\
&+
 (|A_2|^2+|B_2|^2) \biggl[ q^2 \{ (p_+ \cdot k)^2 + (p_- \cdot k)^2 \}   - 2 m_l^2 (q \cdot k)^2 \biggr]   \\
&+ 
2  {\rm Re} ( B^\star_1 A_2 + A^\star_1 B_2 ) q^2 \biggl[ (p_+ \cdot k)^2 - (p_- \cdot k)^2 \biggr]  \biggr\} 
\end{split}
\end{equation}
It is convenient to introduce dimensionless variables 
\begin{equation}
x_\gamma = 2 E_\gamma/M_{B_s} \; , \; x_\pm = 2 E_\pm/M_{B_s} \; , \; \Delta = x_+-x_-  \; , \;  r = m_l^2/M^2_{B_s}
\end{equation}
in terms of which $q^2 = M^2_{B_s} (1-x_\gamma)$. Taking $x_\gamma$ and $\Delta$ as the two coordinates of the Dalitz plot, phase space is defined by
\begin{equation}
\begin{split}
| \Delta | \le v x_\gamma
\; , \;    v & = \sqrt{1-4 m^2_l/q^2} = \sqrt{1-4 r/(1-x_\gamma)} \, ,  \\
0 & \le x_\gamma \le 1- 4 r 
\end{split}
\end{equation}
In terms of $x_\gamma$ and $\Delta$, the differential decay width takes the form 
\begin{equation}\label{diffdecaywidth}
\begin{split}
\frac{d \Gamma}{d x_\gamma d \Delta} 
&= {\cal N} \biggl[ ( |A_1|^2 + |B_1|^2) \biggl\{ \frac{(1-x_\gamma)(x^2_\gamma+\Delta^2)}{8} + \frac{1}{2} r x^2_\gamma \biggr\} \\
&+ (|A_2|^2 + |B_2|^2) \biggl\{ \frac{(1-x_\gamma)(x^2_\gamma+\Delta^2)}{8} - \frac{1}{2} r x^2_\gamma \biggr\}        \\
&+ 2 {\rm Re}(B^\star_1 A_2 + A^\star_1 B_2) (1-x_\gamma) \frac{1}{4} x_\gamma \Delta \biggr]
\end{split}
\end{equation}
where ${\cal N} = \left[ \alpha^2 G^2_F/ (256 \pi^4) \right] | V_{tb} V^\star_{ts} |^2 M^5_{B_s}$. 
The last term is linear in $\Delta$ and produces an asymmetry between the $l^+$ and $l^-$ energy spectra.

We will derive from Eq. (\ref{diffdecaywidth}) two distributions of interest: \\
(i) The charge asymmetry $A(x_\gamma)$ defined as
\begin{eqnarray}\label{asymmetry}
A(x_\gamma) &=& \frac{   \left(  \int_0^{v x_\gamma} \frac{d \Gamma}{d x_\gamma d \Delta} - \int^0_{-v x_\gamma} \frac{d \Gamma}{d x_\gamma d \Delta}  \right) d \Delta }{
\int_{-v x_\gamma}^{+v x_\gamma} \frac{d \Gamma}{d x_\gamma d \Delta} d \Delta } \\
&=& \frac{3}{4} v (1-x_\gamma) \frac{ 2 {\rm Re}(B^\star_1 A_2 + A^\star_1 B_2)}{ \left\{ (|A_1|^2+|B_1|^2)(1-x_\gamma+2r) + (|A_2|^2+|B_2|^2)(1-x_\gamma-4r) \right\} }
\nonumber
\end{eqnarray}
(ii) The photon energy spectrum
\begin{equation}
\frac{d \Gamma}{d x_\gamma} = \frac{\alpha^3 G^2_F}{768 \pi^4} |V_{tb} V^\star_{ts} |^2 M^5_{B_s} v x^3_\gamma \biggl[
(|A_1|^2+|B_1|^2)(1-x+2r) + (|A_2|^2+|B_2|^2)(1-x_\gamma-4r) \biggr]
\end{equation}

To proceed further, we must introduce a model for the form factors which appear in the functions $A_{1,2}$ and $B_{1,2}$ defined in Eq. (\ref{a1etc}).

\section{\bf Model for Form Factors}

First of all, we note that the form factors $f_T$ and $f^\prime_T$ defined in Eq. (\ref{matrixelement}) are necessarily equal, by virtue of the identity
\begin{equation}
\sigma_{\mu \nu} = \frac{i}{2} \epsilon_{\mu \nu \alpha \beta} \sigma^{\alpha \beta} \gamma_5
\end{equation}
This was pointed out by Korchemsky et al \cite{kor}. We therefore have to deal with three independent form factors $f_V, f_A$ and $f_T$. These have been computed in Ref. 
\cite{kor}
using perturbative QCD methods combined with heavy quark effective theory. For the vector and axial vector form factors of the radiative decay $B^+ \to l^+ \nu \gamma$, 
and their tensor counterpart, defined as in Eq. (\ref{matrixelement}), these authors obtain the remarkable result
\begin{equation}
f_V( E_\gamma) = f_A(E_\gamma) = f_T(E_\gamma) = \frac{f_B m_B}{2 E_\gamma} \left( Q_u R - \frac{Q_b}{m_b} \right) \
+ {\cal O} \left( \frac{\Lambda^2_{\rm QCD}}{E^2_\gamma} \right) 
\end{equation}
where $R$ is a parameter related to the light-cone wave-function of the $B$ meson, with an order of magnitude $R^{-1} \sim \bar{\Lambda} = M_B - m_b$, where the
binding energy $\bar{\Lambda}$ is estimated to be between $0.3$ and $0.4$ GeV. Applying the same reasoning to the form factors for $\bar{B}_s \to l^+ l^- \gamma$, 
we conclude that
\begin{eqnarray}
f_V (E_\gamma) = f_A (E_\gamma) = f_T (E_\gamma) = \frac{f_{B_S} M_{B_s}}{2 E_\gamma} \left( -Q_s R_s + \frac{Q_b}{m_b} \right) +
{\cal O} \left( \frac{\Lambda^2_{\rm QCD}}{E^2_\gamma} \right)
\end{eqnarray}
In what follows, we will neglect the term $Q_b/m_b$, and approximate the form factors by
\begin{equation}\label{formfactors}
f_{V,A,T} ( E_\gamma) \approx \frac{f_{B_s} M_{B_s}}{2 E_\gamma} \frac{1}{3 \bar{\Lambda}_s} = \frac{1}{3} \frac{f_B}{\bar{\Lambda}_s} \frac{1}{x_\gamma}
\end{equation}
where $\bar{\Lambda}_s = M_{B_s} - m_b$ will be taken to have the nominal value $0.5$ GeV. Several of our results will depend only on the universal form 
$f_{V,A,T} (E_\gamma) \sim 1/ E_\gamma$, independent of the normalization. As pointed out in \cite{kor}, a check of the behaviour $f_{V,A} \sim 1/E_\gamma$ in the case of 
$B^+ \to \mu^+ \nu \gamma$ is afforded by the photon energy spectrum, which is predicted to be
\begin{equation}
\begin{split}
\frac{d \Gamma}{d x_\gamma} 
& \sim 
\left[ f^2_V ( E_\gamma) + f^2_A (E_\gamma) \right] x^3_\gamma (1-x_\gamma) \\
& \sim  x_\gamma (1-x_\gamma) 
\end{split}
\end{equation}
In the case of the reaction $B_s \to l^+ l^- \gamma$, the normalization of the tensor form factor $f_T (E_\gamma)$ at $E_\gamma = M_B/2 \, ({\rm i.e.} \;  x_\gamma=1)$ 
can be checked by appeal to the decay rate of $B_s \to \gamma \gamma$. To see this connection, we note that the matrix element of 
$B_s \to \gamma(k,\epsilon) + \gamma(k^\prime,\epsilon^\prime)$ can be obtained from that of $B_s \to l^+ l^- \gamma$ by putting $C_9 = C_{10} =0$, and replacing the factor 
$\left( e f_T C_7/q^2 \right) (\bar{l} \gamma_\mu l)$ by $f_T (x_\gamma=1) \epsilon^{\star \prime}_\mu$. This yields the matrix element
\begin{equation}\label{matrixelement2}
\begin{split}
{\cal M} ( \bar{B}_s \to \gamma(\epsilon,k) \gamma( \epsilon^\prime, k^\prime) ) 
&=
 -i \frac{G_F e^2}{\sqrt{2} \pi^2} \left( V_{tb} V^\star_{ts} \right) \cdot
\biggl[ A^+ F_{\mu \nu} F^{\mu \nu \prime} + i A^- F_{\mu \nu} \tilde{F}^{\mu \nu \prime} \biggr] \\
\intertext{with} 
A^+ & = \, - A^- \, = \, \frac{1}{4} \, M_{B_s} \, f_T(x_\gamma=1) \, C_7 \,. 
\end{split}
\end{equation}

The result for $A^\pm$ coincides with that obtained in Refs. \cite{chang,reina,herrlich} when $f_T (x_\gamma=1) = - \frac{Q_d f_B}{\bar{\Lambda}_s} = 
\frac{1}{3} \frac{f_B}{\bar{\Lambda}_s}$. 
( In Ref. \cite{reina,herrlich}, the role of the parameter $\Lambda_s$ is played
by the constituent quark mass $m_s$. ) Thus the decay width of $B_s \to \gamma \gamma$, 
\begin{equation}
\Gamma ( B_s \to \gamma \gamma ) = \frac{M^3_{B_s}}{16 \pi} \left| \frac{G_F e^2}{\sqrt{2} \pi^2} V_{tb} V^\star_{ts} \right|^2
\left( |A_+|^2 + |A_-|^2 \right)
\end{equation}
serves as a test of the normalization factor $f_T ( x_\gamma = 1 )$.

We remark, parenthetically, that the calculation of $B_s \to \gamma \gamma$, based on an effective interaction for $b \to s \gamma \gamma$, produces 
the amplitudes $A^+$
and $A^-$ given in Eq. (\ref{matrixelement2}) in the limit of retaining only the `reducible' diagrams related to the transition $b \to s \gamma$. 
Inclusion of `irreducible' contributions like $b \bar{s} \to c \bar{c} \to \gamma \gamma$ introduces a correction term in $A_-$ causing the ratio $| A_+/A_-|$ to 
deviate from unity. Estimates in Ref. \cite{chang,reina} yield values for this ratio between $0.75$ and $0.9$. The branching ratio $Br( B_s \to \gamma \gamma)$ is estimated at
$5 \times 10^{-7}$, with an uncertainty of about $50 \%$.

Having specified our model for the form factors $f_V(x_\gamma), f_A(x_\gamma)$ and $f_T(x_\gamma)$, we proceed to present results for the spectrum and branching ratio of  
$B_s \to l^+ l^- \gamma$. We use $M_{B_s} = 5.3 \, {\rm GeV}, f_{B_s} = 200 \, {\rm MeV}$ and, where necessary, $\bar{\Lambda}_s = 0.5 \, {\rm GeV}$ 
in the normalization of the form factors in Eq. (\ref{formfactors}).


\section{\bf Results}

\subsection{Charge Asymmetry}

With the assumption of universal form factors $f_V = f_A = f_T \sim \frac{1}{x_\gamma}$, the asymmetry $A(x_\gamma)$ in Eq. (\ref{asymmetry}) assumes the simple form
\begin{equation}
A(x_\gamma) = \frac{3}{4} \, v \, \frac{2 \, C_{10} \, (C_9 + 2C_7 \frac{1}{1-x_\gamma}) \, (1-x_\gamma)}{ (C_9 + 2C_7 \frac{1}{1-x_\gamma})^2 (1-x_\gamma+2r) 
+ C^2_{10} ( 1-x_\gamma-4r) }
\end{equation}
This is plotted in Fig.\ref{plot1}, and is clearly large and negative over most of the $x_\gamma$ domain, changing sign at $x_\gamma = 1 + \frac{2 C_7}{C_9}$.
(A negative asymmetry corresponds to $l^-$ being more energetic, on average, 
than $l^+$ in the decay $\bar{B}_s ( = b \bar{s} ) \to l^+ l^- \gamma.$) The average charge asymmetry is
\begin{equation}
\left< A \right> = \frac{3}{4} \frac{   \int_0^{1-4r} d x_\gamma \, v^2 x_\gamma (1-x_\gamma) 2 C_{10} (C_9  + 2 C_7 \frac{1}{1-x_\gamma})}{    
\int_0^{1-4r} d x_\gamma \, v x_\gamma \left[ (1-x_\gamma+2r)(C_9 + 2 C_7 \frac{1}{1-x_\gamma})^2+(1-x_\gamma-4r) C^2_{10} \right] }
\end{equation}
and has the numerical value $\left< A \right>_e = -0.28$ , 
$\left< A \right>_\mu = -0.47$ for the modes $l=e,\mu$, the difference arising essentially from the end-point region $x_\gamma \approx 1-4 \,r$.

\subsection{Photon Energy Spectrum}

With the form factors of Eq. (\ref{formfactors}), the photon energy spectrum simplifies to
\begin{equation}
\frac{d \Gamma}{d x_\gamma} = \frac{1}{3} {\cal N} v x_\gamma \left\{ (1-x_\gamma + 2r) (C_9 + 2C_7 \frac{1}{1-x_\gamma})^2 + (1-x_\gamma-4r) C^2_{10} \right\}
\end{equation}
where the constant factor ${\cal N}$ is defined after Eq. (\ref{diffdecaywidth}). 
It is expedient to write this distribution in terms of the decay rate of $\bar{B}_s \to \gamma \gamma$. 
We then obtain the prediction
\begin{eqnarray}\label{photonspectrum}
\frac{d \Gamma( \bar{B}_s \to l^+ l^- \gamma)/d x_\gamma}{\Gamma(\bar{B}_s \to \gamma \gamma)} 
&=&
\left\{ \frac{2\alpha}{3 \, \pi} \frac{x^3_\gamma}{(1-x_\gamma)^2} v (1-x_\gamma+2r) \right\} \\
& \cdot &
\left( \frac{1}{x_\gamma} \right)^2 
\left[
\left\{ \eta_9(1-x_\gamma) + 1 \right\}^2 + \left\{ \eta_{10} (1-x_\gamma) \right\}^2 
\frac{1-x_\gamma-4r}{1-x_\gamma+2r} 
\right] \nonumber
\end{eqnarray}
The first factor (in curly brackets \{ \}) is the QED result expected if the decay $\bar{B}_s \to l^+ l^- \gamma$ is interpreted as a Dalitz pair reaction 
$\bar{B}_s \to \gamma \gamma^\star \to \gamma l^+ l^-$, without form factors. The factor $(1/x_\gamma)^2$ results from the universal behaviour $f_{V,A,T} \sim 1/x_\gamma$ 
given in Eq. (\ref{diffdecaywidth}), 
while the last factor is the electroweak effect associated with the coefficients $\eta_9 = C_9/(2 C_7)$ and $\eta_{10} = C_{10}/(2 C_7)$. This 
distribution is plotted in Figs. \ref{plot2} and \ref{plot4}, where the QED result is shown for comparison.

\subsection{Rates and Branching Ratios}

From the photon spectrum given in Eq. (\ref{photonspectrum}), we derive the `conversion ratios'
\begin{equation}
R_l = \frac{    \int_0^{1-4r}   \frac{d \Gamma}{d x_\gamma} ( B_s \to l^+ l^- \gamma)   }{   \Gamma(B_s \to \gamma \gamma)    }
\end{equation}
The numerical values are $R_e = 4.0 \% \text{  and  } R_\mu = 2.3 \%$. These are to be contrasted with the QED result given by
\begin{equation}
R_l (QED) = \frac{2 \alpha}{3 \, \pi} \left[ (1-18 r^2+8 r^3) \ln \frac{1+\sqrt{1-4r}}{1-\sqrt{1-4r}} + \sqrt{1-4r} \, ( -\frac{7}{2} + 13r +4r^2) \right] 
\end{equation}
which yields $R_e (QED)=2.3 \%$, $ R_\mu (QED) = 0.67 \%$. The absolute branching ratios of $\bar{B}_s \to l^+ l^- \gamma$, obtained by taking 
$Br( B_s \to \gamma \gamma) = 5 \times 10^{-7}$ \cite{chang,reina} are $Br( \bar{B}_s \to e^+ e^- \gamma) = 2.0 \times 10^{-8}$, 
$Br( \bar{B}_s \to \mu^+ \mu^- \gamma) = 1.2 \times 10^{-8}$. Our results for the average charge asymmetry $\left< A \right>_l$, the conversion ratios $R_l$ 
and the branching ratios are summarized in Table \ref{table}.


\section{\bf Comments}

(i) The branching ratios calculated by us are somewhat higher than those obtained in previous work \cite{aliev,geng}, 
which used a different parametrization of the form 
factors $f_V, f_A, f_T, f_{T^\prime}$ based on QCD sum rules \cite{aliev} and light-front models \cite{geng}. In particular these parametrizations do not satisfy the relation
$f_T = f^\prime_T$ which, as noted in \cite{kor}, follows from the identity $\sigma_{\mu \nu} = \frac{i}{2} \epsilon_{\mu \nu \alpha \beta} \sigma^{\alpha \beta} \gamma_5$.

\vspace{5mm}
\noindent
(ii) Our predictions for the charge asymmetry $ \left< A \right>$ and the conversion ratio 
$\Gamma ( \bar{B}_s \to l^+ l^- \gamma ) / \Gamma ( \bar{B}_s \to \gamma \gamma )$ are independent of the parameter $\bar{\Lambda}_s$ which appears in the form factor in
Eq. (\ref{formfactors}). The branching ratios in Table \ref{table} assume 
$Br ( \bar{B}_s \to \gamma \gamma ) = 5 \times 10^{-7}$, and can be rescaled when data on this channel are available.

\begin{figure}
  \epsfig{file=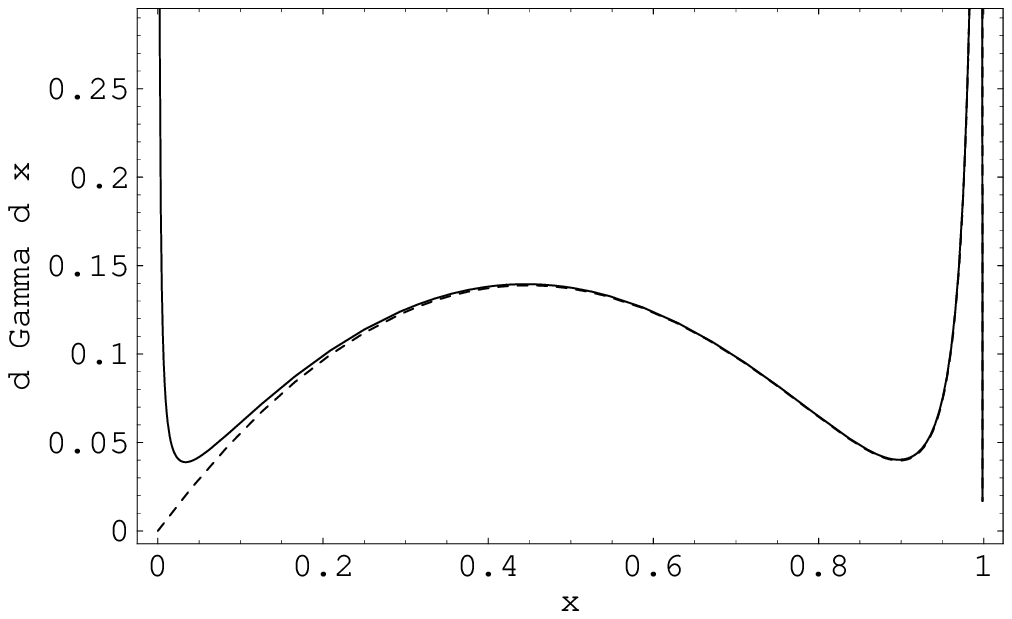}
  \caption{Photon Energy Spectrum in $\bar{B}_s \to \mu^+ \mu^- \gamma$, with bremsstrahlung (solid line) and without bremsstrahlung (dashed line)}
\label{plot6}
\end{figure}

\vspace{5mm}
\noindent
(iii) A full analysis of the decay $\bar{B}_s \to l^+ l^- \gamma$ requires inclusion of the bremsstrahlung amplitude corresponding to photon emission from the leptons in
$B_s \to l^+ l^-$. This contribution is proportional to $f_{B_s} m_l$ and affects the photon energy spectrum in the small $x_\gamma$ region. We have calculated the corrected 
spectrum for $B_s \to l^+ l^- \gamma$, following the procedure in \cite{aliev2}, and the result is shown in Fig. \ref{plot6} for the case $l = \mu$. 
As anticipated, the correction is limited to small $x_\gamma$, and can be removed by a cut at small photon energies.

\begin{figure}
\epsfig{file=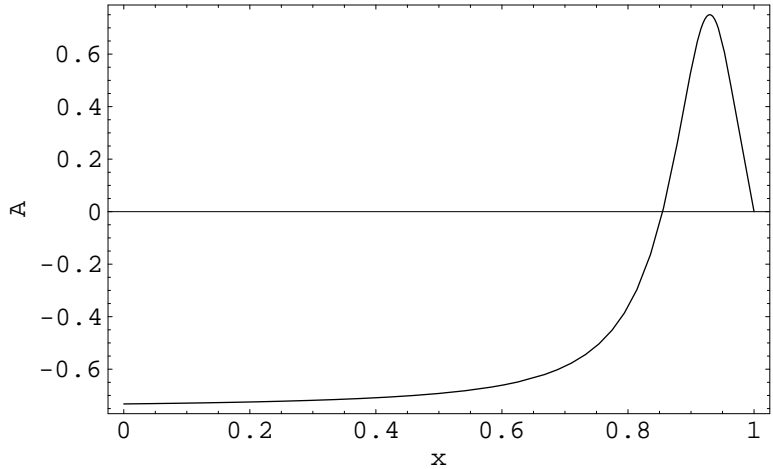}
\caption{Asymmetry versus $x_\gamma$}
\label{plot1}
\end{figure}

\begin{figure}
\epsfig{file=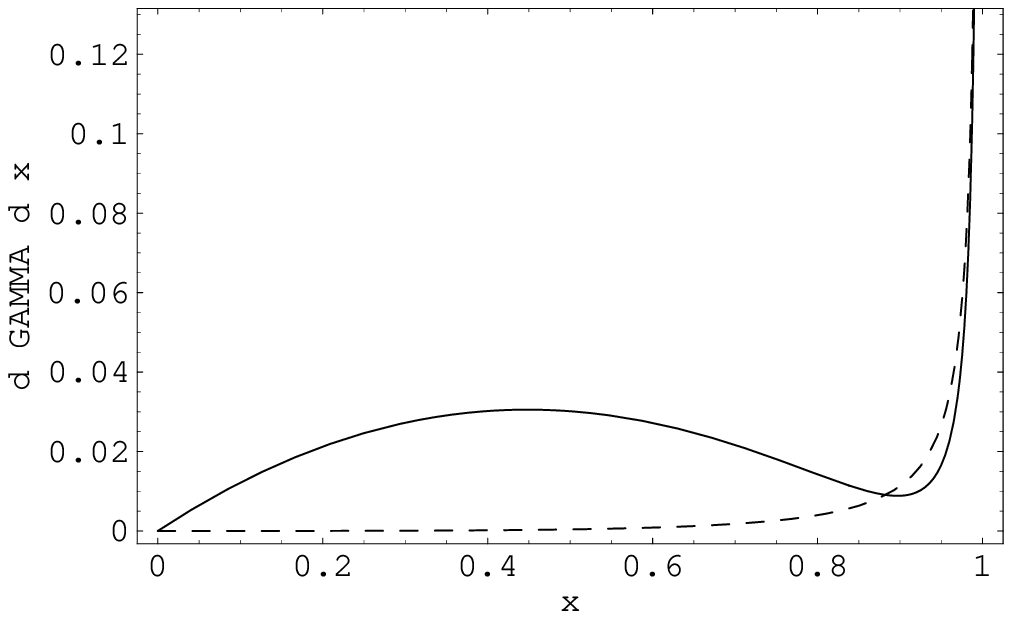} 
\caption{Photon Energy Distribution for $\bar{B}_s \to e^+ e^- \gamma$, normalized to $\bar{B}_s \to \gamma \gamma$. (Dashed line is the QED result.)}
\label{plot2}
\end{figure}

\begin{figure}
\epsfig{file=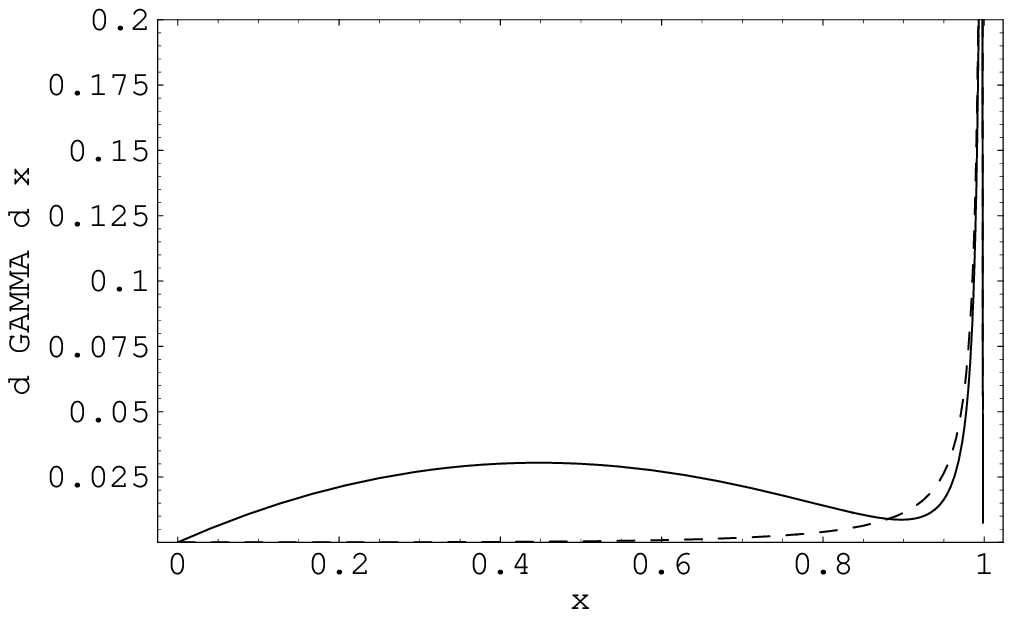}
\caption{Photon Energy Distribution for $\bar{B}_s \to \mu^+ \mu^- \gamma$, normalized to $\bar{B}_s \to \gamma \gamma$. (Dashed line is the QED result).}
\label{plot4}
\end{figure}

\vspace{5mm}
\noindent
(iv) The QCD form factors in Eq. (\ref{formfactors}) 
are valid up to corrections of order $(\Lambda_{\rm QCD}/E_\gamma)^2$. In the small $x_\gamma$ region, arguments based on heavy 
hadron chiral perturbation theory suggest form factors dominated by the $B^\star$ pole with the appropriate quantum numbers, for example,
\begin{equation}
f_V(x_\gamma) \sim \frac{1}{M^2_{B_s} (1-x_\gamma) - M^2_{B^\star_s}} 
\end{equation}
Defining $M_{B^\star_s} - M_{B_s} = \Delta M$, this form factor has the behaviour $f_V (x_\gamma) \sim \frac{1}{x_\gamma + \delta}$, with 
$\delta \approx 2 \Delta M / M_{B_s} \approx 0.02$. We have investigated the effect of replacing the QCD form factor of Eq. (\ref{formfactors}) by a different universal form 
$f_{V,A,T} (x_\gamma) =  f_{B_s} / ( 3 \bar{\Lambda}_s (x_\gamma+\delta) )$, and found only minor changes in the numbers given in Table \ref{table}. 
In general, one must
expect some distortion in the spectrum at low $x_\gamma$, compared to that shown in Figs. \ref{plot1}-\ref{plot6}.

\vspace{5mm}
\noindent
(v) We will examine separately the predictions for $A(x_\gamma)$ and $d \Gamma/d x_\gamma$ in the reaction $B_s \to \tau^+ \tau^- \gamma$, in which the bremsstrahlung
part of the matrix element plays a significant role \cite{aliev2}. We will consider also refinements due to the $q^2-$dependent term in $C_9^{\rm eff}$, and the effects of 
$c \bar{c}$ resonances.

In view of their clear signature, non-negligible branching ratios and interesting dynamics, 
the decays $B_s \to l^+ l^- \gamma$ could form an attractive domain of study at future hadron colliders.

\vspace{3mm}

{\bf Acknowledgments:}
We are indebted to Dr. Gudrun Hiller for drawing our attention to a sign error in the first version of this paper. 
We thank Dr. Thorsten Feldmann for a helpful discussion.
One of us (Y.D.) acknowledges a Doctoral stipend under the Graduiertenf\"orderungs-gesetz of the state of Nordrhein-Westphalen.



\newpage

\large

\begin{table}
\begin{tabular}{|c|c|c|c|} 
\hline
\hline 
 &&& \\
Decay         &  Average Charge        &      Conversion Ratio       &      Branching Ratio        \\
              &   Asymmetry            &                             &                              \\
              &    $\left<   A    \right>$                 &         $\frac{\Gamma( \bar{B}_s \to l^+ l^- \gamma)}{\Gamma(\bar{B}_s \to \gamma \gamma)}$      &      
                           $\frac{\Gamma(\bar{B}_s \to l^+ l^- \gamma)}{\Gamma(\bar{B}_s \to {\rm all})}$          \\
 &&& \\
\hline
 &&& \\
$\bar{B}_s \to e^+ e^- \gamma$     &    $-0.28$      &       $4.0 \%$      &   $2.0 \times 10^{-8}$          \\
 &&&   \\
\hline
   &&&  \\
$\bar{B}_s \to \mu^+ \mu^- \gamma$             &       $-0.47$                &    $2.3 \%$        &        $1.2 \times 10^{-8}$          \\                                 
 &&& \\
\hline
\hline
\end{tabular}
\caption{Average charge asymmetry, Conversion ratio and Branching ratio for the decays 
$\bar{B}_s \to e^+ e^- \gamma$ and $\bar{B}_s \to \mu^+ \mu^- \gamma$. (Last column assumes $Br(\bar{B}_s \to \gamma \gamma) = 5 \times 10^{-7}$)}
\label{table}
\end{table}


\newpage
\vspace{10cm}

\newpage
\vspace{10cm}

\vspace{10cm}
\newpage
\vspace{10cm}

\newpage
\vspace{10cm}


\newpage



\begin{thebibliography}{99}
\bibitem{aliev} T.M. Aliev, A. \"Ozpineci and M. Savci, Phys. Rev. {\bf D55} (1997) 7059
\bibitem{geng} C.Q. Geng, C.C. Lih and W.M. Zhang, Phys. Rev. {\bf D62} (2000) 074017
\bibitem{eilam} G. Eilam, C.D. L\"u and D.X. Zhang, Phys. Lett. {\bf B391} (1997) 461; \\
G. Burdman, T. Goldman and D. Wyler, Phys. Rev. {\bf D51} 9 (1995) 111; \\
G. Eilam, I.Halperin and R.R. Mendel, Phys. Lett. {\bf B361} (1995) 137; \\
P. Colangelo, F.De Fazio, G. Nardulli, Phys. Lett {\bf B372} (1996) 311; \\
D. Atwood, G. Eilam and A. Soni, Mod. Phys. Lett. {\bf A11} (1996) 1061; \\
C.Q. Geng, C.C. Lih and W.M. Zhang, Phys. Rev. {\bf D57} (1998) 5697
\bibitem{kor} G.P. Korchemsky, D. Pirjol and T.M. Yan, Phys. Rev. {\bf D61} (2000) 114510
\bibitem{buch} G. Buchalla, A.J.Buras and M.E. Lautenbacher, Rev. Mod. Phys. {\bf 68} (1996) 1125
\bibitem{xiong} Z. Xiong and J.M. Yang, Nucl. Phys. {\bf B602} (2001) 289
\bibitem{chang} C.V. Chang, G.L. Lin and Y.P. Yao, Phys. Lett. {\bf B415} (1997) 395
\bibitem{reina} L. Reina, G. Ricciardi and A. Soni, Phys. Rev. {\bf D56} (1997) 5805; \\
   G. Hiller and E.O. Iltan, Phys. Lett. {\bf B 409} (1997) 425-437 
\bibitem{herrlich} S. Herrlich and J. Kalinowski, Nucl. Phys. {\bf B381} (1992) 501
\bibitem{kru}F. Kr\"uger and L.M. Sehgal, Phys. Lett. {\bf B380} (1996) 199; \\
C.S. Lim, T. Morozumi and A.I. Sanda, Phys. Lett. {\bf B218} (1989) 343; \\
N.G. Deshpande, J. Trampetic and K.Panose, Phys. Rev. {\bf D39} (1989) 1461; \\
P.J.O'Donnell, M. Sutherland and H.K.K. Tung, Phys. Rev {\bf D46} (1992) 4091
\bibitem{aliev2} T.M. Aliev, N.M. Pak and M. Savci, Phys. Lett. {\bf B424} (1998) 175
\end{thebibliography}
\end{document}